\documentclass{aa}

\usepackage{graphicx}
\usepackage{color}
\usepackage{txfonts}
\begin{document}

\authorrunning{Li et al.}

\title{Heating and cooling of coronal loops observed by SDO}

\author{L. P. Li \inst{1,2}, H. Peter\inst{2}, F. Chen\inst{2}
   and J. Zhang \inst{1}}

\institute{Key Laboratory of Solar Activity, National Astronomical
   Observatories, Chinese Academy of Sciences, 100012 Beijing, China\\
              \email{lepingli@nao.cas.cn}
         \and
             Max Planck Institute for Solar System Research (MPS),
             37077 G\"ottingen, Germany}
\date{Received ...; accepted ...}


\abstract
   {One of the most prominent processes suggested to heat the corona to well above $10^6$\,K builds on nanoflares, short bursts of energy dissipation.}
   {We compare  observations to model predictions to test the validity of the nanoflare process.}
   {Using extreme UV data from AIA/SDO  and HMI/SDO
    line-of-sight magnetograms we study the spatial and temporal evolution of
   a set of loops in active region AR 11850.}
   {We find a transient brightening of loops in emission from \ion{Fe}{xviii} forming at about 7.2\,MK while at the same time these loops dim in emission from lower temperatures. This points to a fast heating of the loop that goes along with evaporation of material that we observe as apparent upward motions in the image sequence. After this initial phases lasting for some 10\,min, the loops  brighten in a sequence of AIA channels showing cooler and cooler plasma, indicating the cooling of the loops over a time scale of about one hour.
A comparison to the predictions from a 1D loop model  shows that
this observation supports the nanoflare process in (almost) all
aspects. In addition, our observations show that the loops get
broader while getting brighter, which cannot be understood in a 1D
model. }
   {}


\maketitle
%

\section{Introduction}

How structures in the upper solar atmosphere, i.e.\ the transition
region and corona, are heated and sustained is one of the major
unresolved issues in solar and stellar astrophysics (e.g.\ Klimchuk
\cite{klim06}). Active regions (ARs) that are dominated  by loops,
prominently seen in extreme ultraviolet (EUV) and X-rays, are the
ideal place to investigate the dominant heating mechanism(s) in the
upper solar atmosphere. The AR loops constitute basic building
blocks and are usually divided into two types: warm loops ($\sim$
1\,MK, Ugarte-Urra et al. \cite{ugar09}) and hot loops ($>$ 2\,MK,
Antiochos et al. \cite{anti03}). These have been vastly studied in
theory and observations for understanding the coronal heating (see a
review in Klimchuk \cite{klim06}).

The processes providing the energy input for the loops can be
categorized into two classes: the steady heating (Reale et al.
\cite{real00}; Antiochos et al. \cite{anti03}; Brooks \& Warren
\cite{broo09}) and impulsive heating (Warren \cite{warr03};
Patsourakos \& Klimchuk \cite{pats06}; Feng \& Gan \cite{feng06};
Tripathi et al. \cite{trip10}). In general, they can be
distinguished by comparing the time scale of the heat input with the
typical coronal cooling time, which is of the order of (a fraction
of) an hour. If there are separate pulses of heating that are
shorter than the cooling time, the heating is considered impulsive.
Steady heating will also be found if the energy input lasts for much
longer than the cooling time (or if many very short pulses come in
rapid succession, so that the corona has no time between the short
pulses to relax). Arguments for both steady and impulsive heating is
found in coronal observations. However, from 3D magnetohydrodynamics
(MHD) modeling this distinction of steady and impulsive heating for
loops is not that clear. Such models show that even on the
\emph{same} fieldline on one leg the heating can be steady, while it
is impulsive on the other leg (Bingert \& Peter
\cite{Bingert+Peter:2011}, Peter \cite{Peter:2015}).

By studying an AR moss area, i.e. footpoints of hot loops, Antiochos
et al. (\cite{anti03}) suggested that the heating is quasi-steady.
Brooks \& Warren (\cite{broo09}) and Dadashi et al. (\cite{dada12})
analyzed the Doppler shifts of an AR moss also finding support for
quasi-steady heating. In contrast, Klimchuk (\cite{klim06}) argued
that most coronal heating mechanisms are impulsive for elemental
magnetic flux strands within a loop. Nanoflares as proposed by
Parker (\cite{Parker:1972,Parker:1988}) are usually considered as
the source of impulsive heating in these strands, and a coronal loop
is considered to be a bundle of unresolved strands (Cargill
\cite{carg94}). Observational arguments have been put forward that
these fundamental strands have to have sizes of 500\,km (Brooks et
al. \cite{Brooks+al:2012}, \cite{Brooks+al:2013}) or even less
(Peter et al. \cite{Peter+al:2013.hic}). Tripathi et al.
(\cite{trip10}) compared the observed and theoretical emission
measure distributions in an AR core, and proposed that the hot loops
are heated by nanoflares. Using imaging in six different channels in
the EUV, Viall \& Klimchuk (\cite{vial12}) analyzed the lightcurves
of coronal loops. By comparing with theoretical models they
suggested that both loops in and surrounding the AR cores are heated
by impulsive nanoflares. By measuring the Doppler shifts in AR moss
in lower temperature lines, Winebarger et al. (\cite{wine13})
provided strong evidence that hot loops are impulsively heated.
Recently, Ugarte-Urra \& Warren (\cite{ugar14}) investigated the
frequency of transient brightenings in an AR core, and found that
there are nearly two to three heating events per hour.

Once the heating ceased, the bundle of loop strands cools down. The
lightcurves of coronal loops  of  channels or spectral lines showing
cooler plasma reach their peaks at progressively later times than
channels showing hotter plasma (Schrijver \cite{schr01}; Warren et
al. \cite{warr02}; M\"uller et al. \cite{mull04}; Peter et al.
\cite{pete12}). This time lag has been interpreted as the result of
hot coronal loop plasma cooling down. Using data from the EUV
Imaging Spectrometer (EIS) on board Hinode, Ugarte-Urra et al.
(\cite{ugar09}) could follow the cooling of loops down to transition
region temperatures after a heat deposition. Viall \& Klimchuk
(\cite{vial12}) observed that there is a time-lag consistent with
cooling plasma not only for loops throughout an AR, but also for the
diffuse emission between the loops. Alissandrakis and Patsourakos
(\cite{alis13}) identified some loops which were initially visible
in the AIA 94\,\AA\ images, subsequently in the AIA 335\,\AA\ and in
one case in the AIA 211\,\AA\ channels, supporting the cooling of
impulsive heated loops. However, with the interpretation of this and
other AIA data sets one has to consider that they are naturally
multi-thermal as they cover a broad range of temperatures (e.g. Del
Zanna et al. \cite{delz11}). Attention should be paid to the
interpretation of the AIA observed features that can be affected by
the contribution of particular spectral lines under certain
conditions (e.g. O'Dwyer et al. \cite{odwy10}).

High-speed evaporative upflows reaching speeds of more than
100\,km\,s$^{-1}$ are predicted by 1D loop models with a prescribed
impulsive heat input as expected for e.g. nanoflares (Antiochos \&
Sturrock \cite{anti79}; Patsourakos \& Klimchuk \cite{pats06}). Such
upflows, albeit at somewhat slower speeds of just below
100\,km\,s$^{-1}$, are also found in a 3D MHD model of an emerging
active regions (Chen et al. \cite{Chen+al:2014}). De Pontieu et al.
(\cite{depo09}) looked at the asymmetry of line profiles and
concluded that the line asymmetry is caused by a high-velocity
upflow at the loop footpoint (Tian et al. \cite{tian11}; Doschek
\cite{dosc12}), which is, however, interpreted by others as
propagating slow magneto-acoustic waves (Gupta et al.
\cite{gupt12}), or being dominated by uncertainties (Tripathi \&
Klimchuk \cite{trip13}). Dadashi et al. (\cite{dada12}) found that
the inner part of a moss area shows blueshift of 5\,km\,s$^{-1}$ for
cooler lines (1.0-1.6\,MK) and 1\,km\,s$^{-1}$ for hotter lines
($\sim$ 2\,MK). Tripathi et al. (\cite{trip12}) presented
observations of upflows in warm loops (0.6-1.6\,MK) with speeds
decreasing with height, and considered them as evidence of
chromospheric evaporative upflows. Orange et al. (\cite{oran13})
studied a catastrophically cooling loop, and observed the plasma
upflows at its footpoint sites at multiple transition region
temperatures. On the other hand, downflows are expected during the
cooling process due to the heated plasma radiatively cooling and
condensing in the loops. Redshifts have been reported previously at
footpoints and along the loop structures supporting the presence of
cooling downflows (Del Zanna \cite{delz08}; Tripathi et al.
\cite{trip09}). Cool plasma sliding down on both sides of coronal
loop with speeds of up to 100\,km\,s$^{-1}$ are also reported by
Schrijver (\cite{schr01}) and has been modeled by M{\"u}ller et al.
(\cite{Mueller+al:2005}). Moreover, Ugarte-Urra et al.
(\cite{ugar09}) detected cooling downflows with velocities in the
range of 40\,km\,s$^{-1}$ to over 105\,km\,s$^{-1}$.

To test the nanoflare model for coronal heating, in particular with
respect to the 1D models of Patsourakos \& Klimchuk (\cite{pats06}),
we choose a set of AR loops observed by the Atmospheric Imaging
Assembly (AIA; Lemen et al. \cite{leme12}) onboard the Solar
Dynamics Observatory (SDO; Pesnell et al. \cite{pesn12}). We
investigate the evolution of heated and cooling loops in detail, and
compare these observations with theoretical models. This shows clear
evidence for nanoflare heating of coronal loops for at least this
set of observed loops.

\section{Observations and data processing}

The AIA instrument consists of  a set of normal incidence EUV
telescopes designed for acquiring solar atmospheric images at ten
wavelength bands. In this study, we use AIA multi-wavelength images
from September 24, 2013 with a time cadence and spatial sampling of
12\,s and 0.6\,\arcsec/pixel to study the evolution of AR loops.
Data from the Helioseismic and Magnetic Imager (HMI; Schou et al.
\cite{scho12}) onboard SDO line-of-sight magnetograms are used  to
investigate the underlying photospheric magnetic field. The spatial
sampling and time cadence of the HMI data are 0.5\,\arcsec/pixel and
45\,s, respectively.

In order to analyze the evolution of the hot plasma in the loops the
contribution of the Fe\,XVIII emission line is isolated from the AIA
94\,\AA\ images using the empirical method devised by Warren et al.
(\cite{warr12}). This is to avoid contamination from the cooler
plasma (mostly around 1\,MK) that also contributes to this channel
(Boerner at al. \cite{Boerner+al:2012}). According to ionization
equilibrium, the Fe\,XVIII line shows plasma around 7.2\,MK.
Following Ugarte-Urra \& Warren (\cite{ugar14}) the Fe\,XVIII images
are obtained from the AIA 94\,\AA\ images by subtracting the
contaminating warm (i.e. around 1\,MK) component to the bandpass.
This warm contribution is computed from a weighted combination of
the emission from the AIA 171\,\AA\ and 193\,\AA\ channels,
respectively dominated by Fe\,X and Fe\,XII emission. This empirical
isolation can be expressed as
\begin{equation}\label{E:FeXVIII}
I_{\rm{Fe\,XVIII}}=I_{94}-A\sum_{i=0}^{3}c_{i}\left(\frac{~f\,I_{171}+(1{-}f)\,I_{193}~}{B}\right)^{\displaystyle i}.
\end{equation}
%
%
Here the weighting is given by
$c=[-7.19\times10^{-2}$,
   $9.75\times10^{-1}$,
   $9.79~\times10^{-2}$,
   $-2.81\times10^{-3}]$,
and $A=0.39$, $B=116.32$, $f=0.31$.
More details are found in Warren et al. (\cite{warr12}) and Ugarte-Urra
\& Warren (\cite{ugar14}).

\section{Heating of loops\label{S:heating}}

On September 24, 2013 the active region NOAA AR 11850 was observed
by SDO at the heliographic position N10\,E20. From 03:00\,UT to
05:00\,UT a set of loops was located to the North end of the AR.
Most importantly, no other loop structures are detected surrounding
these loops, so that it is possible to study an isolated set of
loops. These were heated and subsequently cooled down. First they
showed signatures of brightening
(Sect.\,\ref{S:hot.loop.brightening}) and plasma injection
(Sect.\,\ref{S:hot.loop.motions}) in Fe\,XVIII originating from hot
(heated) plasma together with a dimming in cooler channels
(Sect.\,\ref{S:dimming}). After some time evidence for cooling is
observed (Sect.\,\ref{S:cooling})

\subsection{Brightening of a hot loop in Fe\,XVIII\label{S:hot.loop.brightening}}

\begin{figure*}[!t]
\centering
\includegraphics[width=\textwidth]{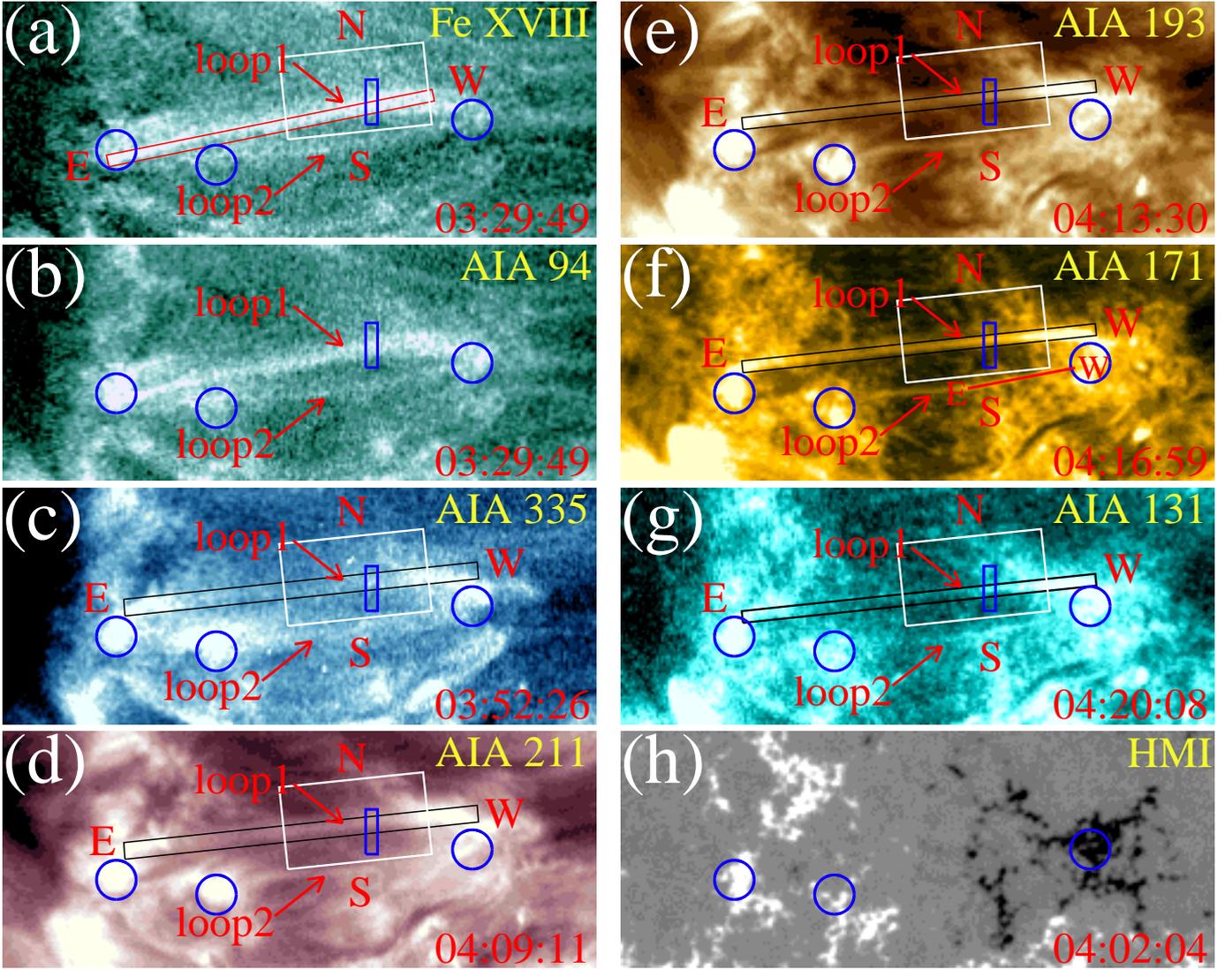}
\caption{AIA/SDO extreme UV images and HMI/SDO magnetogram. Panels
(a-g) display the loops as seen in Fe XVIII (a), AIA 94\,\AA\ (b),
335\,\AA\ (c), 211\,\AA\ (d), 193\,\AA\ (e), 171\,\AA\ (f) and
131\,\AA\ (g). Panel (h) shows the line-of-sight magnetogram. The Fe
XVIII image is derived from the AIA images.\ (see
Eq.\,\ref{E:FeXVIII}). The arrows point to two loops investigated
here. Each of the images (a-g) is shown when the loops become
clearly visible in the respective band. The blue circles mark the
footpoints of the loops, and the blue rectangles in (a-g) the
regions for the lightcurves of the loop1 as shown in
Figs.\,\ref{F:lightcurves} and \ref{F:dimmingcurves}. The white
rectangles NS in (a, c-g) indicate the positions for time-space
diagrams displayed in Fig.\,\ref{F:brightenings}. The red rectangle
EW in (a), the red line EW in (f) and the black rectangles EW in
(c-g) show the positions for space-time diagrams displayed in
Figs.\,\ref{F:heatingmotions}a, \ref{F:heatingmotions}b and
\ref{F:coolingmotions}, respectively. E, W, N and S separately
denote the heliographic directions. The field of view (FOV) is
150$\arcsec$$\times$60$\arcsec$. (An animation of this figure is
available on-line.)}
\label{F:loops}%
\end{figure*}

\begin{figure}
\centering
\includegraphics[width=8.8cm]{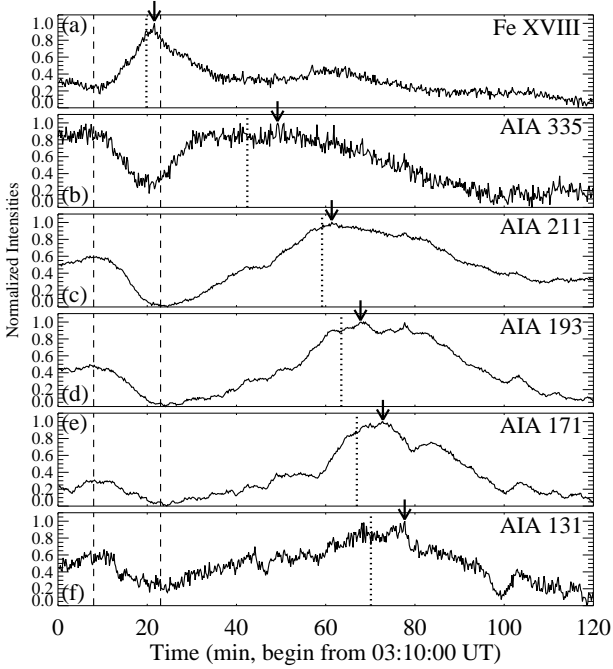}
\caption{Lightcurves of the loop1. The channels are denoted with the
plots. The emission is integrated over  the region marked by the
blue rectangles marked in Fig.\,\ref{F:loops}. The lightcurve in the
AIA 94\,\AA\ channel (not shown) essentially is the same as for the
Fe XVIII line. The dotted lines indicate the times when the loop\,1
became clearly evident in the respective AIA channel as shown in
Figs.\,\ref{F:loops}a-\ref{F:loops}g. The arrows mark the respective
peaks of these lightcurves. The dashed lines indicate the time
interval shown in Fig.\,\ref{F:dimmingcurves}. }
\label{F:lightcurves}%
\end{figure}

Figure\,\ref{F:loops} displays the general information on the loops
in the AIA multi-wavelength observation together with the
information on the magnetic field from HMI. Each coronal band is
shown when the loop\,1 is close to its peak brightness (as indicated
by the dotted lines in  Fig.\,\ref{F:lightcurves} for each of the
bands), spanning roughly one hour. The HMI magnetogram is taken
around the middle of that time interval.  The Fe\,XVIII loop consist
of two components, a northern thick loop, loop\,1, and a southern
thin loop, loop\,2. In this paper, we primarily study the northern
main loop, loop\,1. It first appears in the Fe\,XVIII images at
about 03:20\,UT with a length of nearly 70\,Mm. Moreover, around
this time, there is no corresponding loop visible in other AIA
channels (see movie attached to Fig.\,\ref{F:loops}). On the one
hand this indicates that the loops are heated up the Fe\,XVIII line
characteristic temperature of $\sim$7.2\,MK, because they are not
seen in the cooler channels. On the other hand, the loop is not
heated to temperatures much higher than about 7\,MK, because
otherwise it should be visible in the 131\,\AA\ channel that has a
significant contribution from plasma at around 10\,MK (that is often
seen in flares).

Three blue circles mark the footpoints of the two loops, among which
the western one encircles two neighboring western footpoints of
loop\,1 and loop\,2. To show the relation to the magnetic field, we
overlay them on the line-of-sight magnetogram
(Fig.\,\ref{F:loops}h): the two loops separately connect two
plage-type areas of the AR with opposite polarities.

To determine the overall change in intensity of loop\,1, we
integrate the emission of Fe\,XVIII in the blue rectangle in
Fig.\,\ref{F:loops}a. The resulting (normalized) lightcurve is then
shown in Fig.\,\ref{F:lightcurves}a. (The lightcurve in the AIA
94\,\AA\ channel shows a similar trend). Fe\,XVIII quickly increases
in brightness then decreases again, with the whole brightening of
the loop lasting $\sim$25\,min.

\begin{figure}
\centering
\includegraphics[width=8.8cm]{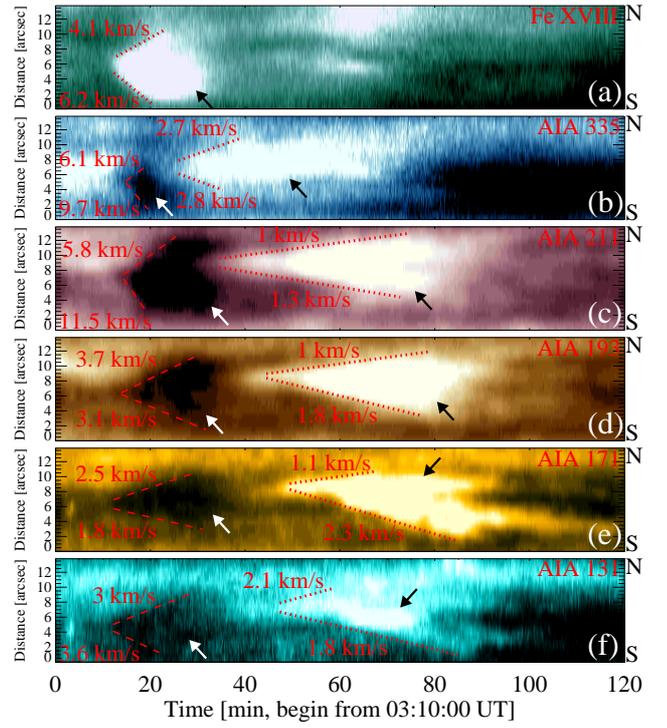}
\caption{Brightening of the loop1. The panels show the time-space
diagrams along the white rectangles NS marked in
Fig.\,\ref{F:loops}. Here the distance is \emph{across} the loop.
The black arrows mark the loop\,1, and the white arrows the dimming.
The dotted lines outline the brightening  expanding perpendicular to
the loop, and the dashed lines the dimming as it expands. The
respective mean velocities are denoted by the numbers in the plots.
N and S are the same as in Fig.\,\ref{F:loops}.}
\label{F:brightenings}%
\end{figure}

This brightening loop is quite thin when it first appears and then
gets broader (i.e. increases its cross section) over the course of
10\,min reaches a width of almost $\sim$8\arcsec\ (this will be
further discussed in Sect.\,\ref{S:loop.broadening}). This is
illustrated in Fig.\,\ref{F:brightenings}, where we show space-time
plots of the evolution of the emission across the loops. For this we
integrate the emission in the white rectangle labeled NS in
Fig.\,\ref{F:loops} in the East-West direction, i.e. along the loop.
For each image of the time series this provides an average variation
of the intensity across the loop. This average variation across the
loop is then plotted in Fig.\,\ref{F:brightenings} as a function of
time for each of the observed channels. Panel a of
Fig.\,\ref{F:brightenings} shows the space-time-plot for Fe\,XVIII.
Here loop\,1 (marked by black arrow) first brightens in the middle,
and then gradually expands to both sides (red dotted lines). The
propagation of the brightening expands with about 5 km\,s$^{-1}$ to
10\,km\,s$^{-1}$ across the loop, respectively.

Comparison with models will have to show, if this motion is a real
motion of the plasma due to an expansion of the loop. Another option
would be that the fieldlines further away from the center of the
loop get heated (and bright) a bit later than the fieldlines in the
center, as one expects from 3D MHD models of loops forming in an
emerging active region (Chen et al.
\cite{Chen+al:2014,Chen+al:2015}).

\subsection{Motions along hot loop in Fe\,XVIII\label{S:hot.loop.motions}}

\begin{figure}[!t]
\centering
\includegraphics[width=7.0cm]{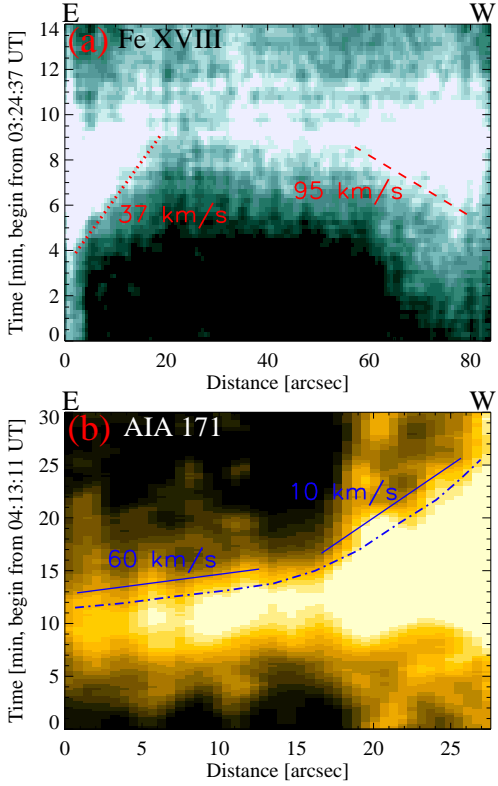}
\caption{Proper motions along the loops. The panels show the
space-time plots of the Fe\,XVIII (a) and AIA 171\,\AA\ (b) images
along the red rectangle EW in Fig.\,\ref{F:loops}a and the red line
EW in Fig.\,\ref{F:loops}f, respectively. In contrast to
Fig.\,\ref{F:brightenings} here the distance is \emph{along} the
loop. The red dotted and dashed lines in (a) and the blue
dash-dotted and solid lines in (b) indicate the proper motions. The
respective mean speeds are denoted by the numbers in the plots. E
and W are the same as in Fig.\,\ref{F:loops}.}
\label{F:heatingmotions}%
\end{figure}

To investigate the motions along the loops we create time-space
plots similar to the ones above, but now showing the (average)
variation along the loop (roughly in the East-West direction). For
this we integrate the intensity (in the red rectangle in
Fig.\,\ref{F:loops}a) along the N-S direction and then plot this
average versus time. This is shown for Fe\,XVIII  along the loop1 in
Fig.\,\ref{F:heatingmotions}a.

At both footpoints we see upward proper motions along the loop. The
upflow at both feet start at almost at the same time
($\sim$03:28\,UT) and move with about 40\,km\,s$^{-1}$ and almost
100\,km\,s$^{-1}$ at the Eastern and Western footpoints,
respectively. This proper motion could be a propagation front of
enhanced emission due to increasing temperature, or/and a signature
of an actual evaporative plasma flow into the loop in response to
increased heating.

\subsection{Dimming in cooler channels\label{S:dimming}}

Almost simultaneous with the appearance of the Fe\,XVIII loops, a
dimming takes place in the AIA channels imaging cooler plasma
($<$7\,MK) as is most clear in  the lightcurves of
Fig.\,\ref{F:lightcurves}. This dimming is also clearly visible in
the images and difference images as shown in Fig.\,\ref{F:dimmings}.
The overlay of the brightening in Fe\,XVIII as a contour on top of
the base difference images (Figs.\,\ref{F:dimmings} d-f) makes clear
that the dimming in the cool channels is not only co-temporal with
the brightening in Fe\,XVIII showing the hot plasma, but that it is
also co-spatial.

\begin{figure}[!t]
\centering
\includegraphics[width=8.0cm]{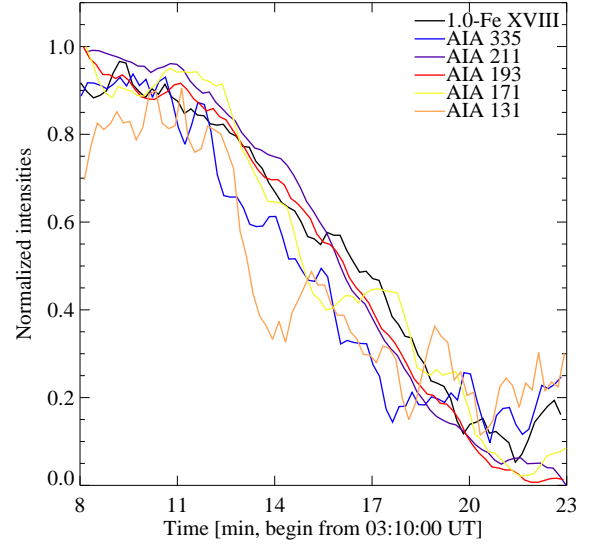}
\caption{Similar to Fig.\,\ref{F:lightcurves}, but for the time
range between two dashed lines as marked in
Fig.\,\ref{F:lightcurves}. All the lightcurves are normalized. In
the case of \ion{Fe}{xviii} we show the inverse of the lightcurve to
make clear the cool channels get dark synchronous with
\ion{Fe}{xviii} getting bright.}
\label{F:dimmingcurves}%
\end{figure}

\begin{figure*}
\sidecaption
\includegraphics[width=12cm]{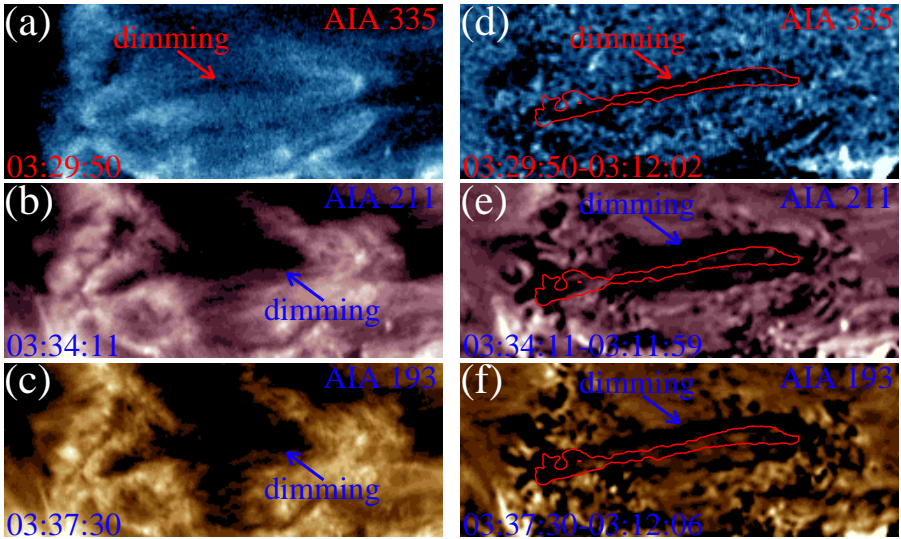}
\caption{AIA/SDO extreme UV images and corresponding base-difference
images.  Panels (a-c) show the original images, panels (d-f) the
respective difference images (d-f). The red contours in (d-f) mark
the Fe\,XVIII loops in Fig.\,\ref{F:loops}a. Same FOV as in
Fig.\,\ref{F:loops}. }
   \label{F:dimmings}%
\end{figure*}

To highlight the very close timing of the brightening of the hot and
the dimming of the cool plasma we overplot normalized curves for the
loop brightness in Fig.\,\ref{F:dimmingcurves}. Here the cool
channels (335\,\AA, 211\,\AA, 193\,\AA, 171\,\AA\ and 131\,\AA) are
all normalized to the average intensity before the event. For the
hot plasma we plot  $I=1.0-I_{\rm{Fe\,XVIII}}$, where
$I_{\rm{Fe\,XVIII}}$ is the Fe\,XVIII normalized to before the
event. All the lightcurves have the same shape falling remarkably
well on top of each other. This underlines
 that the dimming is really co-temporal with the
brightening of the Fe\,XVIII loop.

Similarly as for Fe\,XVIII we check a space-time-plot to investigate
the evolution of the dimming in the cool channels across the loop
(Figs.\,\ref{F:brightenings} b-f). Similar to the brightening in
Fe\,XVIII, all the cool channels show a consistent expansion across
the loop with a speed of the order of 5\,km\,s$^{-1}$ to
10\,km\,s$^{-1}$ that is co-spatial and co-temporal.

From this we conclude that the dimming of the cool channels is
indeed exactly correlated with the brightening of the hot plasma.
This provides strong observational support that the loop under
investigation was quickly heated to some 7\,MK.

\section{Cooling of loops\label{S:cooling}}

After the brightening in the Fe\,XVIII line in response to heating
(Sect.\,\ref{S:heating}), the loop subsequently brightens in bands
showing cooler plasma (Sect.\,\ref{S:cool.loop.brightening}) and
shows signs of plasma draining (Sect.\,\ref{S:cool.loop.draining})
in response to cooling.

\subsection{Loop brightening in cooler channels\label{S:cool.loop.brightening}}

Figures\,\ref{F:loops}c-\ref{F:loops}g show the loop subsequently in
the cooler channels of AIA. We show the snapshots at times  when the
loops are clearly visible and mark these times by the dotted lines
in the lightcurves shown in
Figs.\,\ref{F:lightcurves}b-\ref{F:lightcurves}f. Similar to the
Fe\,XVIII loops shown in Fig.\,\ref{F:loops}a, the loops in the
cooler channels also consist of two sub-loops, loop\,1 and loop\,2.
Their footpoints are located at the same positions as those of the
Fe\,XVIII loops, indicated by the blue circles in
Figs.\,\ref{F:loops}c-\ref{F:loops}g. Furthermore, the loops are
\emph{not} detected in AIA 304\,\AA\ channel, which shows emission
at around 0.1\,MK. This is  cooler than for the AIA 131\,\AA\
channel, suggesting that the loops are cooling down to, but not much
lower than,  $\sim$0.6\,MK which is temperature of maximum
contribution of the AIA 131\,\AA\ channel.

Figure\,\ref{F:lightcurves}b-\ref{F:lightcurves}f displays the
lightcurves of loop\,1 in the cooler  channels of AIA. They show
that the loop begins to brighten after the initial dimming. The
arrows in Fig.\,\ref{F:lightcurves} indicate the peak times of the
loop intensities in the respective channels. The gradual cooling is
clearly discernible through the peak times of the brightening
according to the temperature of maximum contribution. The beginning
and peak times of the AIA channels are listed in
Table\,\ref{tables:peakintensities}. One can easily identify the
time delay of the loop brightening in different wavelength bands
during the cooling process.

Figures\,\ref{F:brightenings}b-\ref{F:brightenings}f illustrate the
evolution of  the brightening of loop\,1 in the cooler channels
after time $\sim$30\,min (indicated by the  black arrows). These
show space-time plots, where the variation across the loop is
averaged along the respective white rectangles shown in
Fig.\,\ref{F:loops}.  Similar to the initial brightening of the
Fe\,XVIII loop in Fig.\,\ref{F:brightenings}a,  in the cooler
channels the loops begin to brighten also in the middle, and then
this brightening propagates to both sides, as indicated by the
dotted lines in Figs.\,\ref{F:brightenings}b-\ref{F:brightenings}f
(see further discussion in Sect.\,\ref{S:loop.broadening}). Here the
propagation is in the range of 1\,km\,s$^{-1}$ to 3\,km\,s$^{-1}$,
i.e. smaller than in the case of the Fe\,XVIII loop. The gradual
appearance of the loop is slightly different in the different
wavelength channels. Moreover, the loop appears about 1\arcsec\ to
the north of the Fe\,XVIII loop for the AIA 335\,\AA\ channel, and
about 2\arcsec\ for other AIA cooler channels. This offset most
likely is due to a slow ($\sim$1 km\,s$^{-1}$) transversal motion of
the loop. This is also supported by the offset between the dimming
(in e.g. 193\,\AA) and the later brightening in the same channel
showing a similar offset. Still, errors due to co-alignment of the
images (using the standard SolarSoft package,
http://www.lmsal.com/solarsoft/) might not be negligible. In
conclusion, the brightening of the cooler loops is attributed to the
cooling after a short initial phase of heating the loops to at least
7\,MK.

\subsection{Motions along cooler loops\label{S:cool.loop.draining}}

\begin{figure*}
\sidecaption
\includegraphics[width=12cm]{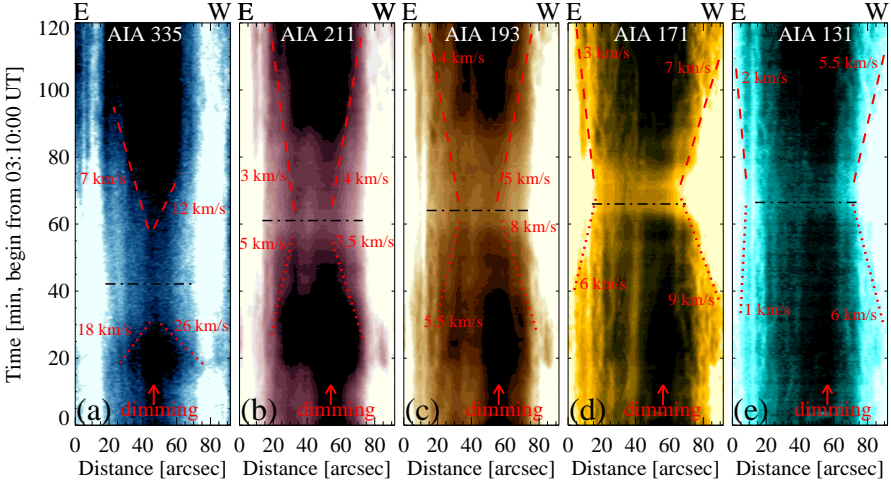}
\caption{Proper motions along the loop. Similar to
Fig.\,\ref{F:heatingmotions}, but along the black rectangles EW in
loop\,1 marked in Figs.\,\ref{F:loops}c-\ref{F:loops}g. The red
dotted and dashed lines indicate upward and downward proper motions
near the loop footpoints. The respective mean velocities are denoted
by the numbers in the plots. The black dash-dotted lines indicate
when the loop becomes first visible in the respective channel (same
times as in Fig.\,\ref{F:lightcurves}).  E and W are the same as in
Fig.\,\ref{F:loops}. }
\label{F:coolingmotions}%
\end{figure*}

To study the (proper) motions along the loops we investigate
space-time plots with the spatial dimension roughly aligned with the
loops. Here we restrict the analysis to the cool channels because we
concentrate on the cooling phase of the loops. For this we show the
temporal evolution along the black rectangles EW (see
Figs.\,\ref{F:loops}c-\ref{F:loops}g) in
Fig.\,\ref{F:coolingmotions} in the form of space-time plots. These
basically show the evolution in the cool channels along loop\,1.
This also underlines that the cool channels show a dimming while the
Fe\,XVIII loop gets bright (around time 20 min; cf.
Sect.\,\ref{S:dimming} and Fig.\,\ref{F:lightcurves}).

After the dimming, upward proper motions along the loop from the
footpoints to the apex are seen in the space-time plots in
Fig.\,\ref{F:coolingmotions} (after time 20\,min, indicated by
dotted red lines). The average speeds are higher for the AIA
335\,\AA\ channel with the values of $\sim$20 km\,s$^{-1}$, and
smaller for other cooler channels with the values of several
km\,s$^{-1}$. The dash-dotted lines in Fig.\,\ref{F:coolingmotions}
marks the appearances of the loop after the upward proper motions.
From this is also clear that the loop appears in a gradual fashion.
After the loop appeared, downward proper motions along the loop from
the apex to the footpoints are observed, indicated by the dashed
lines in Fig.\,\ref{F:coolingmotions}. The mean speeds are several
km\,s$^{-1}$ for all these AIA channels.

The above discussion concentrates on loop\,1 (cf.
Fig.\,\ref{F:loops}). We also checked the proper motions along
loop\,2. There, downward motions from the apex to the western
footpoint are detected for all the cooler AIA channels.
Figure\,\ref{F:heatingmotions}b shows an example displaying a
space-time plot of a series of AIA 171\,\AA\ images along the red
line EW as shown in Fig.\,\ref{F:loops}f. The dash-dotted line marks
this downward motion, with an average speed of 60 km\,s$^{-1}$ at
the beginning. It then decreases to 10\,km\,s$^{-1}$, denoted by two
blue solid lines.

\begin{table*}[!ht]
\caption{Properties of the AIA/SDO multi-wavelength loop and comparison to the nanoflare model.}
\label{tables:peakintensities}
{
\begin{tabular}{ccccccc|cccc}         
\hline\hline              

         & temperature      &  \multicolumn{2}{c}{obs. brightening}&  &          && density            & AIA temperature      &                     &
\\
\cline{3-4}
 AIA     & of max. contrib. & start& peak & \multicolumn{2}{c}{cooling time lag} && in model           & response kernel      & \multicolumn{2}{c}{peak intensity in AIA}
\\
\cline{5-6} \cline{10-11}
channel  & $T_{\rm{max}}$   & time & time &       observation    &    model      && at $T_{\rm{max}}$: & $K_i(T_{\rm{max}})$  &    model   &  observation
\\
$[$\AA$]$& [MK]             & [UT] & [UT] &          [min]       &    [min]      && $n$ [cm$^{-3}$]    & [cm$^{5}$DN/pixel/s] & $DN$/pixel &  $0.45\,{\times}\,DN$/pixel
\\
\hline

94  & 7.2 & 03:20 & 03:32 &  0 &  0 && 1.0$\times$10$^{9}$ & 4.0$\times$10$^{-27}$ & 2.4 & 7.2  \\

335 & 2.5 & 03:30 & 03:59 & 27 & 29 && 1.4$\times$10$^{9}$ & 5.0$\times$10$^{-27}$ & 5.9 & 6.0  \\

211 & 1.9 & 03:34 & 04:11 & 39 & 40 && 1.2$\times$10$^{9}$ & 1.9$\times$10$^{-25}$ & 164 & 191  \\

193 & 1.5 & 03:34 & 04:18 & 46 & 47 && 1.0$\times$10$^{9}$ & 5.5$\times$10$^{-25}$ & 330 & 329  \\

171 & 0.9 & 03:34 & 04:23 & 51 & 61 && 0.7$\times$10$^{9}$ & 1.2$\times$10$^{-24}$ & 353 & 320  \\

131 & 0.6 & 03:34 & 04:28 & 56 & 68 && 0.6$\times$10$^{9}$ & 6.0$\times$10$^{-26}$ &  13 &  15  \\
\hline

\end{tabular}
}
\tablefoot{
The timing of the observed brightening (start and peak time) is derived from the lightcurves in Fig.\,\ref{F:lightcurves} (see Sect.\,\ref{S:cool.loop.brightening}). The (cooling) time lag following from this in observations and the time lag in the model is discussed in detail in item (4) in Sect.\,\ref{S:comp.prop}.
\\
The four right columns, i.e., model density $n$, Kernel $K_i$ and the modeled and observed peak intensities are discussed in detail in Sect.\,\ref{S:quant.comp}.
}
\end{table*}

\section{Comparison to a nanoflare model}

\subsection{Summary of a nanoflare model}

The behaviour of the loop(s) studied in here using observations by
AIA matches very well to the 1D model of a loop being heated through
a nanoflare  by Patsourakos \& Klimchuk (\cite{pats06}). In their
base model they prescribe the (volumetric) energy input as being
uniformly distributed along a 150\,Mm long loop at constant cross
section. Their initial loop (in equilibrium) reaches a peak
temperature of 2.5\,MK. They then increase the heat input by a
factor of 33 (from 0.03\,W/m$^3$ to  1\,W/m$^3$) for 250\,s. This
leads to a transient temperature increase and evaporation of
chromospheric material into the loop followed by a cooling and
draining phase after the heating ceased. In their model the increase
of the heating is simply prescribed. However, such variations of the
energy input are also found in 3D MHD models where the heating and
dynamics is driven self-consistently by stressing the magnetic field
through horizontal convective motions in the photosphere (Bingert \&
Peter \cite{Bingert+Peter:2011,Bingert+Peter:2013}). The major
properties of the loop heated by the nanoflare in base model  of
Patsourakos \& Klimchuk (\cite{pats06}) can be summarized as follows
(based mostly on their Fig.\,1 which shows the properties ``halfway
up the leg of the strand for the base model''):
\begin{enumerate}

\item Following the heat pulse the plasma heats from 2.5\,MK to about 7.5\,MK. This temperature rises quickly over the course of less than 5 minutes.

\item The loop would be clearly visible over some 12 minutes during and after the nanoflare heating pulse in emission from hot plasma (their estimate is based on  \ion{Fe}{xvii} being brighter than half of the peak intensity).

\item The upflows of hot plasma associated with the chromospheric evaporation reach some 60\,km\,s$^{-1}$ around the time when the loop is at its maximum temperature. The maximum upflow speeds (closer to the footpoints) can reach up to 200\,km\,s$^{-1}$.

\item After the heat pulse the plasma needs some 40 minutes to cool down to about 2\,MK. The cooling continues and after some 65 minutes the temperature dropped below 1\,MK (this is the end of the evolution shown in their paper).

\item In the cooling phase there is a slow net downflow of the order of 10 km/s that is gradually emptying the loops. After the heat pulse and the initial evaporation the density remains fairly constant for almost half an hour and then starts decreasing by about a factor of two for the next half hour.

\end{enumerate}

\subsection{Comparison to model properties}\label{S:comp.prop}

These properties of the model match the case of the observed loop we
present in this study very well. In the following we compare these
model properties to the observations as discussed in
Sects.\,\ref{S:heating} and \ref{S:cooling} item by item.

(1)~~ The loop is seen first in the AIA channels representing
temperatures up to some 2\,MK (i.e. the 211\,\AA\ channel). Within a
few minutes these cooler channels darken and the \ion{Fe}{xviii}
gets more intense, consistent with a temperature rise from some
2\,MK to above 7\,MK (cf. Figs.\,\ref{F:lightcurves} and
\ref{F:dimmingcurves}). The identical spatial, temporal and
kinematical relationships between the Fe\,XVIII loops and the cooler
dimming suggest that the dimming is indeed attributed to a quick
rise of the loop temperatures. The temperature rise of the coronal
plasma (Zhang et al. \cite{zhang12}) in the loops and their
surrounding atmosphere after the impulsive heating leads to the
brightening in the AIA higher temperature channel, e.g. 94\,\AA, and
the dimming in the AIA lower temperature channels. While Patsourakos
\& Klimchuk (\cite{pats06}) show the temperature only at ``halfway
up the leg'', the efficient heat conduction at these high
temperatures will lead to a relatively flat temperature profile, so
that the temperature variation will be comparable all along in the
coronal part of the loop. Because we do not see a brightening in the
131\,\AA\ channel, which has a contribution also at 10\,MK we can
conclude that the loop we observe says below 10\,MK, just as in the
model.

(2)~~ The brightening in \ion{Fe}{xviii} derived from the AIA
94\,\AA\ channel lasts for some 11 minutes (cf.\
Fig.\,\ref{F:lightcurves}a) when applying the criterion as
Patsourakos \& Klimchuk (\cite{pats06}) in their model. We observed
\ion{Fe}{xviii} forming at about 7\,MK, Patsourakos \& Klimchuk
(\cite{pats06}) synthesized \ion{Fe}{xvii} forming at 5\,MK.
However, the two contribution functions overlap quite a bit, so that
the two lines can be expected to behave (very roughly) similar. In
$\log{T}$ the difference in formation temperature is about 0.15,
while the full-width-at-half-maximum for \ion{Fe}{xvii} and
\ion{Fe}{xviii} is about 0.4 and 0.4, respectively (according to
Chianti v7.1.3, Landi et al. \cite{Landi+al:2013}).

(3)~~ In our observations we find for the first time a high-speed
upflow of \emph{hot} plasma at 7\,MK filling a loop. Here we find
speeds of some 40\,km\,s$^{-1}$ to almost 100\,km\,s$^{-1}$ at the
legs of the loop (see Fig.\,\ref{F:heatingmotions}). This is
consistent with the nanoflare model by Patsourakos \& Klimchuk
(\cite{pats06}). Previously reported upflow speeds mostly have been
slower and always have been at significantly lower temperatures. For
instance, Tripathi et al. (\cite{trip12}) found speeds of
4\,km\,s$^{-1}$ to 10\,km\,s$^{-1}$ at  temperatures of 0.6\,MK to
1.6\,MK, Orange et al. (\cite{oran13}) saw speeds below
10\,km\,s$^{-1}$ at below 1\,MK, De Pontieu et al. (\cite{depo09})
found upflows of 50\,km\,s$^{-1}$ to 100\,km\,s$^{-1}$ at below
2\,MK, all of which would not be consistent with the Patsourakos \&
Klimchuk (\cite{pats06}) model, where the upflows are seen in the
hot plasma.

(4)~~ Qualitatively the cooling time in the observation from the
brightening in hot \ion{Fe}{xviii} line to the enhancement in the
171\,\AA\ channel is about one hour (Figs.\,\ref{F:loops} and
\ref{F:lightcurves}), and thus matching the nanoflare model. For a
more quantitative estimate of the timing of the cooling in the
observation we evaluate the time lag between the peak of the
emission in the AIA\ channels with respect to the peak in the
94\,\AA\ channel (see Table\,\ref{tables:peakintensities} and
Fig.\,\ref{F:lightcurves}). This can be compared to the time lag
when the temperature in the model loop matches the temperature of
maximum contribution, $T_{\rm{max}}$, which is listed in
Table\,\ref{tables:peakintensities}, too. The model and the observed
time lags match quite well.

(5)~~ In the later part of the cooling phase we see an apparent
downward motion of the emission in the cool channels. Mostly these
are of the order of 4\,km\,s$^{-1}$ to 12\,km\,s$^{-1}$ (dashed
lines in Fig.\,\ref{F:coolingmotions}). These would be similar to
the downflows during the cooling and draining phase in the model of
Patsourakos \& Klimchuk (\cite{pats06}) and the observations of Del
Zanna (\cite{delz08}) and Doschek et al. (\cite{dosc08}). However,
they are much smaller than previously reported values of about
100\,km\,s$^{-1}$ (Schrijver \cite{schr01}), 60\,km\,s$^{-1}$
(Tripathi et al. \cite{trip09}), or 39\,km\,s$^{-1}$ to
105\,km\,s$^{-1}$(Ugarte-Urra et al. \cite{ugar09}). In the earlier
part of the cooling phase, we also see upflows in the cool channels
ranging from  5\,km\,s$^{-1}$ to 25\,km\,s$^{-1}$ (dotted lines in
Fig.\,\ref{F:coolingmotions}), which would not be consistent with
the nanoflare model. Such upflows have also been reported by Orange
et al. (\cite{oran13}) who interpreted these as the result of
magnetic reconnection at the loop footpoints. However, it might well
be that the apparent motions of the emission as seen in
Fig.\,\ref{F:coolingmotions} are \emph{not} real mass motion, but
just a cooling front moving along the loop. This process  has been
reported by Peter et al. (\cite{Peter+al:2006}) who found an
apparent upward motion of coronal emission even if the actual plasma
flow is downward (their Fig.\,6 and attached movie). Thus the
\emph{apparent} downward motion of the emission might be consistent
with the nanoflare model of Patsourakos \& Klimchuk (\cite{pats06}).

\subsection{Quantitative comparison to nanoflare model}\label{S:quant.comp}

Because the imaging data do cannot provide reliable information on
the plasma density, we use the coronal emission to be expected from
the nanoflare model of  Patsourakos \& Klimchuk (\cite{pats06}) for
a quantitative comparison with the observations. For this we derive
the count rate, or digital number
\begin{equation}\label{E:count.rate}
DN_i=K_i(T)\,n^2\,\ell\,t
\end{equation}
from the temperature $T$ and number density $n$ of the model. For
the temperature response of the AIA channels we use the kernels
$K_i(T)$ discussed by Boerner et al. (\cite{Boerner+al:2012}) as
provided through the SolarSoft package and the Chianti atomic
database package (v7.1.3, Landi et al. \cite{Landi+al:2013}). We
assume that the length of the line-of-sight $\ell=3$\,Mm through the
loops is comparable to the  diameter of the loop as found in our
observations. The exposure time $t=2$\,s is set to the typical value
used in the AIA extreme UV wavebands.

For the quantitative comparison we estimate the peak emission in
each of the AIA bands. While in principle one could use the original
data to construct full lightcurves (as done e.g. by Peter et al.
\cite{pete12}), for our estimate it should be sufficient to take the
density $n$ from the model at the time the temperature of the model
equals the temperature of maximum contribution, $T_{\rm{max}}$ (see
Table\,\ref{tables:peakintensities}). These values are taken from
the base model of  Patsourakos \& Klimchuk (\cite{pats06}) shown in
their Fig.\,1. We then evaluate Eq.\,(\ref{E:count.rate}) for
$T{=}T_{\rm{max}}$ and list the density $n$ at $T_{\rm{max}}$ and
the kernel values $K_i(T_{\rm{max}})$ together with the resulting
count rates $DN_i$ for the model at peak emission in channel $i$ in
Table\,\ref{tables:peakintensities}.

To compare these values to our observations, we determine the peak
intensity halfway up the loops in the blue rectangles shown in
Fig.\,\ref{F:loops}, because the data shown by Patsourakos \&
Klimchuk (\cite{pats06}) in their Fig.\,1 are also taken there. To
avoid background effects, we subtract the emission before the
brightening (in the cool channels during the dimming minimum). These
values we multiply by 0.45 and list them in
Table\,\ref{tables:peakintensities}. When multiplying by this factor
the model values of the cool channels match quite well to the
observations. Because the real loops have a variation of intensity
across the loop, the 1D model loop will have to represent some
average loop with all properties being constant perpendicular to the
loop axis. Thus it seems reasonable that the peak values in the
observations should be reduced somewhat, here by about a factor of
2, to match the average model values. In contrast to the cool
channel, the hot 94\,\AA\ channel shows a significantly lower
emission in the model. Considering the simplistic approach taken
here to derive the model values, it might well be that for a  proper
synthesis of the 94\,\AA\ emission one gets higher values, in
particular when considering the fast temporal evolution during the
heating phase with fast flows filling the loop leading to
considerable deviations of ionization equilibrium (which is
implicitly assumed when using the kernels $K_i(T)$ to estimate the
AIA\ emission).

Summarizing this comparison of count rates and the  discussion in
Sect.\,\ref{S:comp.prop} we conclude that the observations of the
loop we presented here match the nanoflare model of Patsourakos \&
Klimchuk (\cite{pats06}) very well. Therefore this observation can
be considered as a confirmation of this nanoflare-heating concept,
at least for some loops in the solar corona.

\subsection{Broadening of the loops}\label{S:loop.broadening}

An interesting feature of the loop we investigate here is its
expansion perpendicular to the loop axis --- the loop gets thicker
in time! This is illustrated in Fig.\,\ref{F:brightenings} and has
been mentioned already in Sects.\,\ref{S:hot.loop.brightening} and
\ref{S:cool.loop.brightening}. In the heating phase as well as
during the cooling phase the loops first appear to be bright in the
middle and then seem to get wider, i.e. they increase their cross
section with an apparent speed of 1\,km\,s$^{-1}$  to
5\,km\,s$^{-1}$. This also applies to the dimming of the cool AIA
channels in the heating phase.

To explain this behaviour one could speculate that the magnetic
field expands in response to the heating of the loop. However, this
expansion is seen in the heating and in the cooling phase in total
over the course of almost one hour. With the typical expansion
speeds this would correspond to an increase of the cross section of
some 10\,Mm. Because the initial loop is thinner than 1\,Mm, this
would go along with a significant reduction of the magnetic field
strength, according to flux conservation by a factor of 100! This
seems quite unlikely. Thus think that one can rule out  the
possibility that this expansion of the loop seen in AIA emission is
due to the expansion of the magnetic tube hosting the loop.

An alternative explanation would be that the fieldlines in the
magnetic tube get heated stronger at its center than further away
from the center, and consequently the latter fieldlines will get
brighter later. One would arrive at a scenario where the center of
the magnetic tube gets heated stronger and thus gets bright in the
hot 94\,\AA\ channel before the fieldlines away from the center of
the tube get bright. It remains to be seen why then  the center of
the loop also seems to cool earlier than the fieldlines away from
the center. Maybe simply the whole time evolution of the fieldlines
away from the center is delayed in time, which would explain why the
center is again seen earlier in the cooling phase in the cooler AIA
channels. The speed of which the loop expands in the transverse
direction is different in the heating and in the cooling phase (cf.
Fig. 3). This might be because the heating being impulsive and the
cooling being a slower process. However, modeling might reveal the
importance of other processes, such as thermal conduction.

As far as we know, this peculiar behaviour has not been reported
before, and certainly this increase of the width of the loop in time
cannot be explained by a 1D model. 3D MHD models predict that the
structure in loops seen in emission might mismatch the structure of
the magnetic field which might explain the constant cross section of
coronal loops (Peter \& Bingert \cite{Peter+Bingert:2012}).
Likewise, the dynamic evolution of the emission might decouple from
the evolution of the magnetic field (Chen et al.
\cite{Chen+al:2015}). This indicates that the phenomenon we see here
is a 3D effect of different fieldlines in a magnetic tube hosting a
coronal loop getting heated at different times. However, detailed
(3D) models of this are needed to draw conclusions on this
behaviour.

\section{Summary}

Using AIA/SDO multi-wavelength images and HMI/SDO context
magnetograms, we investigate the heating and cooling processes of a
set of loops to the north of the AR 11850 on September 24, 2013. The
loops were heated up to $\sim$7.2\,MK as observed in the Fe\,XVIII
emission. Simultaneously, a dimming surrounding the Fe\,XVIII loops
took place in the cooler AIA cooler channels, reported here for the
first time. After the Fe\,XVIII loop appears, upward motions along
the main loop are detected from both  footpoints at  high
temperature (7\,MK). Afterwards, the loop cools and  appears in
sequence in the cooler AIA. This and other key observables match
very well to a 1D  model simulating nanoflare heating of a loop by
Patsourakos \& Klimchuk (\cite{pats06}). The sequence of the
brightening of the loop first in the hot and then in the
progressively cooler channels matches the model not only
qualitatively, but also quantitatively in terms of the timing and
even of the count rates observed on the Sun and estimated from the
model. The very good match of observation and model indicates that
this loop might constitute the first detailed case of an individual
loop being heated by a nanoflare.

However, open questions remain. In particular the significant
broadening of the loop during the heating and the cooling phase
cannot be explained by a 1D model. It might be due to the different
fieldlines in the magnetic tube hosting the loop being heated at
different times and/or with different power. A conclusive answer can
only be achieved by further observations and new (3D) models of the
heating of coronal loops.

\begin{acknowledgements}

This work is supported by NASA contract NNG09FA40C (IRIS), the
Lockheed Martin Independent Research Program, the European Research
Council grant agreement No. 291058 and NASA grant NNX11AO98G.
The AIA and HMI data used are provided courtesy of NASA/SDO and the
AIA and HMI science teams.
This work is supported by the National Basic Research Program of
China under grant 2011CB811403, the National Natural Science
Foundations of China (11303050, 11533008, 11025315, 11221063,
11322329, 11303049), the CAS Project KJCX2-EW-T07, and Young
Researcher Grant of National Astronomical Observatory, Chinese
Academy of Sciences.
F. C. was supported by the International Max-Planck Research School
(IMPRS) for Solar System Science at the University of G\"ottingen.
\end{acknowledgements}


\begin{thebibliography}{}


\bibitem[2013]{alis13} Alissandrakis, C. E., \& Patsoourakos, S.
2013, A\&A, 556, A79

\bibitem[1979]{anti79} Antiochos, S. K., \& Krall, K. R. 1979, ApJ,
229, p.\,788

\bibitem[2003]{anti03} Antiochos, S. K., et al. 2003, ApJ, 590, p.\,547

\bibitem[2011]{Bingert+Peter:2011} Bingert, S., \& Peter, H. 2011,
A\&A, 530, A112

\bibitem[2013]{Bingert+Peter:2013} Bingert, S., \& Peter, H.
2013, A\&A, 550, A30

\bibitem[2012]{Boerner+al:2012} Boerner, P. F., Edwards, C. G., Lemen, J. R.,
et al. 2012
Sol.\ Phys., 275, 41

\bibitem[2009]{broo09} Brooks, D. H., \& Warren, H. P. 2009, ApJ,
703, p.\,10

\bibitem[2012]{Brooks+al:2012} Brooks, D. H., Warren, H. P., \& Ugarte-Urra,
I.
2012, ApJ, 755, L33

\bibitem[2013]{Brooks+al:2013} Brooks, D. H., Warren, H. P., Ugarte-Urra,
I.,
\& Winebarger, A. R. 2013, 772, L19

\bibitem[1994]{carg94} Cargill, P. J. 1994, ApJ, 422, p.\,381

\bibitem[2014]{Chen+al:2014} Chen, F., Peter, H., Bingert, S., \& Cheung,
M. C. M. 2014, A\&A, 564, A12

\bibitem[2015]{Chen+al:2015} Chen, F., Peter, H., Bingert, S., \& Cheung,
M. C. M. 2015, Nature Phys., 11, 492

\bibitem[2012]{dada12} Dadashi, N., Teriaca, L., Tripathi, D.,
Solanki, S., \& Wiegelmann, T. 2012, A\&A, 548, A115

\bibitem[2009]{depo09} De Pontieu, B., McIntosh, S. W., Hansteen, V.
H., \& Schrijver, C. 2009, ApJ, 701, L1

\bibitem[2008]{delz08} Del Zanna, G. 2008, A\&A, 481, L49

\bibitem[2011]{delz11} Del Zanna, G., O'Dwyer, B., \& Mason, H. E.
2011, A\&A, 535, A46

\bibitem[2008]{dosc08} Doschek, G. A., Warren, H. P., Mariska, J.
T., et al. 2008, ApJ, 686, p.\,1362

\bibitem[2012]{dosc12} Doschek, G. A. 2012, ApJ, 754, p.\,153

\bibitem[2006]{feng06} Feng, L., \& Gan, W. Q. 2006, Chinese Journal of
Astronomy and Astrophysics, 6, p.\,608

\bibitem[2012]{gupt12} Gupta, G. R., Teriaca, L., Marsch, E.,
Solanki, S. K., \& Banerjee, D. 2012, A\&A, 546, A93

\bibitem[2006]{klim06} Klimchuk, J. A. 2006, Sol. Phys., 234, 41

\bibitem[2013]{Landi+al:2013} Landi, E., Young, P. R., Dere, K. P., et al. 2013, ApJ, 763, 86

\bibitem[2012]{leme12} Lemen, J., Title, A., Akin, D., et al. 2012,
Sol. Phys., 275, p.\,17

\bibitem[2004]{mull04} M\"uller, D., Peter, H., \& Hansteen, V.
2004, A\&A, 424, 289

\bibitem[2005]{Mueller+al:2005} M{\"u}ller, D. A. N., De Groof, A., Hansteen,
V. H. \& Peter, H. 2005, A\&A, 436, 1067

\bibitem[2010]{odwy10} O'Dwyer, B., Del Zanna, G., Mason, H. E.,
Weber, M. A., \& Tripathi, D. 2010, A\&A, 521, A21

\bibitem[2013]{oran13} Orange, N., Chesny, D., Oluseyi, H., et al.
2013, ApJ, 778, p.\,90

\bibitem[1972]{Parker:1972} Parker, E. N. 1972,
ApJ, 174, 499

\bibitem[1988]{Parker:1988} Parker, E. N. 1988,
ApJ, 330, 474

\bibitem[2006]{pats06} Patsourakos, S., \& Klimchuk, J. A. 2006, ApJ, 647,
p.\,1452

\bibitem[2012]{pesn12} Pesnell, W., Thompson, B., \& Chamberlin, P.
2012, Sol. Phys., 275, p.\,3

\bibitem[2015]{Peter:2015} Peter, H. 2015,
Phil.\ Trans.\ R.\ Soc.\ A, 373, 20150055

\bibitem[2012]{Peter+Bingert:2012} Peter, H., Bingert, S.
2012, A\&A, 548, A1

\bibitem[2012]{pete12} Peter, H., Bingert, S., \& Kamio, S. 2012,
A\&A, 537, A152

\bibitem[2013]{Peter+al:2013.hic} Peter, H., Bingert, S., Klimchuk, J. A., et al. 2013,
A\&A, 556, A104

\bibitem[2006]{Peter+al:2006} Peter, H., Gudiksen, B. V., Nordlund, \AA.
2006, ApJ, 638, 1086

\bibitem[2000]{real00} Reale, F., Peres, G., Serio, S., et al. 2000,
ApJ, 535, p.\,423

\bibitem[2012]{scho12} Schou, J., Scherrer, P., Bush, R., et al.
2012, Sol. Phys., 275, p.\,229

\bibitem[2001]{schr01} Schrijver, C. J. 2001, Sol. Phys., 198,
p.\,325

\bibitem[2011]{tian11} Tian, H., McIntosh, S. W., \& De Pontieu, B.
2011, ApJ, 727, L37

\bibitem[2009]{trip09} Tripathi, D., Mason, H., Dwivedi, B., Del
Zanna, G., \& Young, P. R. 2009, ApJ, 694, p.\,1256

\bibitem[2010]{trip10} Tripathi, D., Mason, H., \& Klimchuk, J.
2010, ApJ, 723, p.\,713

\bibitem[2012]{trip12} Tripathi, D., Mason, H., Del Zanna, G., \&
Bradshaw, S. 2012, ApJ, 754, L4

\bibitem[2013]{trip13} Tripathi, D., \& Klimchuk, J. A. 2013, ApJ,
779, p.\,1

\bibitem[2009]{ugar09} Ugarte-Urra, I., Warren, H., \& Brooks, D.
2009, ApJ, 695, p.\,642

\bibitem[2014]{ugar14} Ugarte-Urra, I., \& Warren, H. 2014, ApJ,
783, p.\,12

\bibitem[2012]{vial12} Viall, N. M., \& Klimchuk, J. A. 2012, ApJ,
753, p.\,35

\bibitem[2002]{warr02} Warren, H., Winebarger, A., \& Hamilton, P.
2002, ApJ, 579, L41

\bibitem[2003]{warr03} Warren, H., Winebarger, A., \& Mariska,
J. 2003, ApJ, 593, p.\,1174

\bibitem[2012]{warr12} Warren, H., Winebarger, A., \& Brooks, D.
2012, ApJ, 759, p.\,141

\bibitem[2013]{wine13} Winebarger, A., Tripathi, D., Mason, H. E.,
\& Del Zanna, G. 2013, ApJ, 767, p.\,107

\bibitem[2012]{zhang12} Zhang, J., Yang, S. H., Liu, Y., \& Sun, X.
D. 2012, ApJ, 760, L29

\end{thebibliography}
\end{document}